# Power and temperature dependent model for High Q superconductor resonators

A. Alexander[1, 2], C.G. Weddle[1], C.J.K. Richardson[1, 3]

[1] Laboratory for Physical Sciences, University of Maryland, 8050 Greenmead Dr., College Park MD 20740, USA

[2] Department of Physics, University of Maryland, College Park, Maryland 20742, USA

[3] Department of Material Science and Engineering, University of Maryland, College Park, Maryland 20742, USA

## Abstract

Measuring the internal quality factor of coplanar waveguide superconducting resonators is an established method of determining small losses in superconducting devices. Often, resonator losses are attributed to two-level system (TLS) defects using a power dependent model for the quality factor. However, excess non-equilibrium quasiparticles can also limit the quality factor of superconducting resonators. Here, a two-temperature, power and temperature dependent model is used to evaluate resonator losses for titanium nitride and aluminum coplanar waveguide resonators with loss tangents of $10^{-5}$ to $10^{-6}$. This model enables a set of parameters to describe both the internal quality factor and fractional frequency shift of the resonator across a broad range of average cavity photon numbers, from 1 to $10^{6}$, and temperatures between those limited by the experimental apparatus near 50 mK and thermal quasiparticle dominated. The presented model also offers potential insight into observing the temperature and power dependence of TLS-TLS interactions in these resonators.

## I. INTRODUCTION

Low-loss superconducting circuit elements are critical for high coherence superconducting quantum circuits. Coplanar waveguide (CPW) resonators are an essential component for qubit devices and are used as readout elements in two-dimensional transmon[1–3] and flux[4,5] qubits. CPW resonators also provide value for material characterization[6–9].

Superconducting quantum circuits suffer from microwave loss originating from two-level system (TLS) defects[10,11], nonequilibrium quasiparticles[12–14], vortices[15,16], microwave radiation[17], and coupling with package modes[18]. These loss mechanisms have a characteristic dependence on applied microwave power, bath temperature, magnetic field, frequency of operation, and resonator geometry.

Conventionally, the power dependence of the internal quality factor, $Q_i$, of CPW resonators is measured at the lowest possible bath temperature. This measurement approach is mostly suited for devices that are limited by TLS loss, and observation of high power and low power saturation enables quantification of the TLS loss. The TLS loss can be difficult to assess if loss saturation at high powers is not observed[19]. Unfortunately, other loss mechanisms are difficult to isolate from power-dependent measurements at a single temperature. Of particular concern is the loss contribution from non-equilibrium quasiparticles that is difficult to quantify[20] but may be comparable to TLS losses for some refined resonator materials and fabrication processes.

While TLS loss is relatively straightforward to calculate, modeling non-equilibrium quasiparticle loss is calculated by solving coupled nonlinear energy-dependent rate equations of phonon and quasiparticle systems[20–23]. For the fully self-consistent nonlinear model, the energy-dependent equations require several poorly known or unobservable material-dependent parameters. Additionally, these models are most often applied in high-power regimes where quasiparticle losses are expected to dominate instead of the high-quality material and low-power regime that is relevant to quantum information devices.

This article describes a two-temperature and power dependent model that simultaneously considers TLS and non-equilibrium quasiparticle losses in a superconducting resonator. The TLS loss is calculated using the model expressed in terms of the lifetime and decoherence time of an interacting TLS ensemble[24]. The quasiparticle loss is calculated by representing the quasiparticle number density with an effective temperature that may differ from the bath temperature. This approximation is suitable to describe thermalized, non-equilibrium quasiparticle behavior[25]. Additionally, a single electron relaxation rate is included in the quasiparticle rate equations to provide an energy pathway to restore equilibrium between the effective quasiparticle and bath temperatures without recombination. Temperature and power dependent loss measurements of high-quality factor aluminum and titanium nitride thin film resonators are used to illustrate the effectiveness of this model to identify the TLS and quasiparticle contributions to the measured microwave loss.

## II. RESONATOR MODEL

### A. Two-Level Systems

At low temperatures and low powers, unsaturated TLS defects resonantly interact with the microwave electric field in the waveguide resonator. This interaction leads to energy transfer to the phonon population, resulting in losses. For the resonators studied here that operate in the 5 GHz range, the resonant energy is on the order of 20 μeV. This energy is small compared to the accuracy of current numerical methods that might be used for the simulation and identification of the precise microscopic structure of TLS defects[26]. However, it is known to be associated with dielectrics adjacent to or near superconducting materials such as oxides[11], hydroxyl groups[27], or chemical residues[28]. In coplanar waveguide resonators grown on low-loss substrates, TLS defects primarily reside at the metal-air, metal-substrate, and substrate-air interfaces[29].

The degree of saturation of an interacting ensemble of TLS defects determines the dielectric loss. The loss is maximum at low power and low temperature when the TLSs are primarily in the ground state and decreases with either increasing temperature or power when the upper state TLS occupancy increases thus reducing the ability of the TLS ensemble to resonantly absorb microwave photons. The temperature and power dependent loss, $Q_{TLS}$, of an ensemble of TLSs including the TLS–TLS interaction, can be expressed as[24]

$$\frac{1}{Q_{TLS}(\omega_r, n, T)} = \frac{1}{Q_{TLS}^0}\left(\frac{\tanh\frac{\hbar\omega_r}{2k_BT}}{\sqrt{1+\left(\frac{n^{\beta_2}}{DT^{\beta_1}}\right)\tanh\frac{\hbar\omega_r}{2k_BT}}}\right). \tag{1}$$

The contribution of the TLS defects to the dielectric constant is reflected by the change in the resonance frequency. The temperature dependent fractional resonance frequency shift is given by[10,11,24]

$$\frac{\delta f(\omega_r, T)}{f_r} = \frac{1}{\pi Q_{TLS}^0}\left[\Re\left(\Psi\left(\frac{1}{2}+\frac{1}{2\pi i}\frac{\hbar\omega_r}{k_BT}\right)\right) - \log\frac{\hbar\omega_r}{2\pi k_BT}\right]. \tag{2}$$

In these equations, the TLS loss coefficient, $(Q_{TLS}^0)^{-1} = \sum_i p_i \tan\delta_I$, is determined from the sum of the loss tangent, $\tan\delta_i$, and participation ratio, $p_i$, of the different regions. In Eq. (2) $\Re(\Psi(x))$ is the real part of the complex digamma function evaluated at x. For coplanar waveguide resonators, the individual contributions from different dielectric materials such as the native oxide at the metal-air interface and substrate-air interface, dielectrics at the metal-substrate interface, or in the substrate are considered. Conventional terms of TLS behavior are the geometric resonance frequency of the resonator, $f_r = \omega_r/(2\pi)$, temperature, T, the average number of microwave photons in the cavity, n, and $\beta_2$ defines the strength of the saturation. In lieu of the single critical photon number, we follow the expression presented in ref [24] $n_c \propto 1/T_{1_{TLS}}T_{2_{TLS}}$, where the average lifetime of an interacting TLS ensemble is expressed as $T_{1_{TLS}} \propto \tanh(\hbar\omega_r/2k_BT)$, and the decoherence time is $T_{2_{TLS}} \propto \left(DT^{\beta_1}\right)^{-1}$. The variables associated with describing the decoherence of the TLS ensemble, D and $\beta_1$, are empirical parameters and unitless to satisfy dimensional analysis.

### B. Quasiparticles

In superconductors, quasiparticles provide a resistive channel. According to the BCS theory[30] the thermal quasiparticle density vanishes at low temperatures compared to the critical temperature of the superconductor. This is an important motivation for operating superconducting qubits at low-millikelvin temperatures despite the common usage of materials that have a critical temperature, $T_c$, around 1K. However, there are experimental observations of limited performance of superconducting quantum devices that are attributed to excess quasiparticles despite sample measurement temperatures near 10 mK (1% of $T_c$of aluminum)[31–34].

Theoretically, any photon or phonon with energy greater than the superconducting gap (2Δ) can break Cooper pairs and generate quasiparticles. These pair-breaking particles may originate from stray infrared, microwave, or optical photons[12], cosmic or gamma rays[13], spurious antenna modes[14], or non-thermalized input signals having excess high-energy photons[35]. Even sequential multiphoton absorption of sub-gap energy microwave signals can lead to quasiparticle excitation. Their single-particle relaxation from an excited energy state to the band edge can generate sufficiently high energy phonons to break Cooper pairs[20,21].

Calculating the energy averaged number density of net quasiparticles, $n_{qp}$, can be represented as a two-temperature rate equation,

$$\frac{\partial \mathrm{n_{qp}}(\mathrm{T_{qp}})}{\partial \mathrm{t}} = \mathrm{I_{ext}} + \Gamma_{\mathrm{G}} \mathrm{N_{2\Delta}}(\mathrm{T_b}) - \Gamma_{\mathrm{R}} \mathrm{n_{qp}^2}(\mathrm{T_{qp}}) - \Gamma_{\mathrm{S}} \left( \mathrm{n_{qp}}(\mathrm{T_{qp}}) - \mathrm{n_{qp}}(\mathrm{T_b}) \right), \tag{3}$$

with terms that represent external excitation, $\mathrm{I_{ext}}$, quasiparticle generation through Cooper pair breaking by high energy phonons, $\mathrm{N_{2\Delta}}$, quasiparticle relaxation into Cooper pairs, and relaxation of excited quasiparticles towards thermal equilibrium. Here, $\Gamma_{\mathrm{G}}$, is the pair braking rate, $\Gamma_{\mathrm{R}}$, is the quasiparticle recombination rate, and $\Gamma_{\mathrm{S}}$, is the single quasiparticle relaxation rate that returns the quasiparticle to thermal equilibrium. The two temperatures, $\mathrm{T_b}$ and $\mathrm{T_{qp}}$, represent the bath temperature and effective quasiparticle temperature.

The key assumption of the effective temperature model in eq. 3 is that the population density of each particle-type is adequately described by the appropriate thermal distribution such that the complete microscopic model[21,36] does not need to be used. This assumption is valid when the intraparticle scattering rate is larger than the interparticle scattering rate. The effective temperature model allows time-varying non-equilibrium conditions to be described, but it is not suitable for strong continuous excitations that cause spectral hole burning or other athermal population distributions, or pulsed dynamics that are on a timescale of the intraparticle scattering rates.

Under conditions that the superconductor achieves thermal equilibrium, there is no external energy coupled into the quasiparticle population and the quasiparticles are in thermal equilibrium with phonons, i.e., $I_{ext} = 0$ and $T_{qp} = T_b$. The quasiparticle and the 2Δ phonon number densities simplify to

$$\Gamma_{\mathrm{G}} N_{2\Delta}(T_b) = \Gamma_R n_{qp}^2(T_b). \tag{4}$$

However, when equilibrium conditions are not met, such as when quasiparticles are excited by a modest external energy source ($I_{ext} \neq 0$) the quasiparticles may no longer be in thermal equilibrium with the bath and more accurately described by an effective quasiparticle temperature that is above the bath temperature ($T_{qp} > T_b$). The external energy may be black body energy coupled into the superconducting resonator through the rf lines, or a leaky sample enclosure. In the presence of a constant (continuous wave) external energy source, the steady-state relationship between the quasiparticle and the pair-breaking phonon number densities becomes

$$\Gamma_{\mathrm{R}} \left( \mathrm{n_{qp}^2}(\mathrm{T_{qp}}) - \mathrm{n_{qp}^2}(\mathrm{T_b}) \right) + \Gamma_{\mathrm{S}} \left( \mathrm{n_{qp}}(\mathrm{T_{qp}}) - \mathrm{n_{qp}}(\mathrm{T_b}) \right) = \mathrm{I_{ext}}. \tag{5}$$

The positive root is the quasiparticle number density with an elevated quasiparticle temperature ($T_{qp} > T_b$),

$$\mathrm{n_{qp}}(\mathrm{T_{qp}}) = \sqrt{\frac{\mathrm{I_{ext}}}{\Gamma_{\mathrm{R}}} + \left( \mathrm{n_{qp}}(\mathrm{T_b}) + \frac{\Gamma_{\mathrm{S}}}{2\Gamma_{\mathrm{R}}} \right)^2} - \frac{\Gamma_{\mathrm{S}}}{2\Gamma_{\mathrm{R}}}. \tag{6}$$

The Fermi Dirac distribution function in the low particle density limit, $\mathrm{f}(\mathrm{E}, \mathrm{T_{qp}}) = \exp(-\mathrm{E}/\mathrm{k_B T^*})$ where E is the quasiparticle excitation energy, uses the effective temperature, $\mathrm{T^*}$, to describe either the non-equilibrium quasiparticle temperature, $\mathrm{T_{qp}}$, or equilibrium quasiparticle temperature, $\mathrm{T_b}$. A thermalized quasiparticle number density[37], can be expressed as,

$$\mathrm{n_{qp}}(\mathrm{T^*}) = 2\mathrm{N}(0)\sqrt{2\pi \mathrm{k_B T^*} \Delta(\mathrm{T^*})} \exp\left( \frac{\Delta(\mathrm{T^*})}{\mathrm{k_B T^*}} \right). \tag{7}$$

Here $N(0)$ is the normal state single-spin density of states at the Fermi level and $k_B$ is Boltzmann's constant. For a superconductor with a critical temperature, $T_c$, at low-temperatures ($T^*/T_c < 0.3$) the temperature dependent superconducting gap from BCS theory is approximated as[38]

$$\Delta(T^*) \approx \Delta(0)\exp\left[-\sqrt{\frac{2\pi k_B T^*}{\Delta(0)}}\exp\left(-\frac{\Delta(0)}{k_B T^*}\right)\right]. \tag{8}$$

Where the superconducting gap at T= 0 K is $\Delta(0) \approx 1.76\, k_B T_C$,. When $T^* = T_b$ the quasiparticle number density represents thermally generated quasiparticles, and when $T^* = T_{qp} \neq T_b$ the quasiparticle number density represents a thermalized distribution of quasiparticles that is not in equilibrium with the bath. This situation can occur when energy is coupled into the electron population, and the electron-electron interaction is much stronger than the electron-phonon interaction. The changes in the quasiparticle density impact the resistivity and the kinetic inductance of the superconductor, which affects the resonator through the CPW loss and frequency shift. The quasiparticle contribution to the rf loss is[39].

$$\frac{1}{Q_{QP}(\omega_r, T_{qp})} = -\alpha_{L_{ki}}\gamma\frac{\sigma_1(\omega_r, T_{qp})}{\sigma_2(\omega_r, 0)}. \tag{10}$$

Similarly, the fractional resonance frequency shift is given by

$$\frac{\delta f(T_{qp})}{f_r} = -\frac{\alpha_{L_{ki}}}{2}\gamma\frac{\delta\sigma_2(\omega_r, T_{qp})}{\sigma_2(\omega_r, 0)}. \tag{11}$$

Here $\alpha_{L_{ki}}$ is the kinetic inductance fraction that is frequency, material, and geometry dependent; and $\omega_r = 2\pi f_r$ is the resonator frequency. The characteristic superconductor length scales are captured in the usual manner through $\gamma$ that takes the values of either -1, -1/2, or -1/3 depending on the superconductor film condition[39]. The 100-nm-thick TiN film studied here satisfies the thin film condition so $\gamma_{TiN}$ = -1, while the 100-nm-thick Al film falls in the extreme anomalous limit so $\gamma_{Al}$ = -1/3. The frequency-dependent complex conductivity, $\sigma(\omega_r) = \sigma_1(\omega_r) - i\sigma_2(\omega_r)$, depends on the quasiparticle distribution function[40]. The specific quasiparticle effective temperature is solved numerically using Eq. (7) to connect the quasiparticle rate equations to the real and imaginary components of the complex conductivity for $\omega < 2\Delta/\hbar$ [39]. The complex conductivity relationships are

$$\frac{\sigma_1(\omega, T^*)}{\sigma_n} = \frac{4\Delta(T^*)}{\hbar\omega}\exp\left(-\frac{\Delta(T^*)}{k_B T^*}\right)\sinh\left(\frac{\hbar\omega}{2k_B T^*}\right)K_0\left(\frac{\hbar\omega}{2k_B T^*}\right), \tag{12}$$

$$\frac{\sigma_2(\omega, T^*)}{\sigma_n} = \frac{\pi\Delta(T^*)}{\hbar\omega}\left[1 - \sqrt{\frac{2\pi k_B T^*}{\Delta(T^*)}}\exp\left(-\frac{\Delta(T^*)}{k_B T^*}\right) - 2\exp\left(-\frac{2\Delta(T^*) + \hbar\omega}{2k_B T^*}\right)I_0\left(\frac{\hbar\omega}{2k_B T^*}\right)\right]. \tag{13}$$

In the above equations, $\sigma_n$ is the normal state conductivity before the superconducting transition, $I_0$ is a zeroth-order modified Bessel function of the first kind, and $K_0$ is the zeroth-order modified Bessel function of the second kind. Additional details about the various material parameters are in the supplemental material.

### C. Complete Model

The complete comprehensive temperature and power dependent model can be defined by combining all the different loss mechanisms,

$$\frac{1}{Q_i(\omega_r, T_b, n)} = \frac{1}{Q_A} + \frac{1}{Q_{TLS}^0}\left(\frac{\tanh\frac{\hbar\omega}{2k_B T_b}}{\sqrt{1 + \left(\frac{n^{\beta_2}}{DT_b^{\beta_1}}\right)\tanh\frac{\hbar\omega}{2k_B T_b}}}\right) - \gamma\alpha_{L_{ki}}\frac{\sigma_1(\omega_r, T_{qp})}{\sigma_2(\omega_r, 0)}. \tag{14}$$

All the temperature and power independent loss mechanisms are grouped in the constant first term $1/Q_A$. The second term describes the TLS loss, and the third describes the quasiparticle loss. Similarly, the fractional frequency shift can be defined by combining TLS and quasiparticle contributions,

$$\frac{\delta f(\omega_r, T_b)}{f_r} = \frac{1}{\pi Q^0_{TLS}}\left[\mathrm{Re}\left(\Psi\left(\frac{1}{2}+\frac{1}{2\pi i}\frac{\hbar\omega_r}{k_B T_b}\right)\right) - \log\frac{\hbar\omega_r}{2\pi k_B T_b}\right] - \gamma\frac{\alpha_{L_{ki}}}{2}\frac{\delta\sigma_2(\omega_r, T_{qp})}{\sigma_2(\omega_r, 0)}. \quad (15)$$

It is important to note that the resonance frequency, $f_r$, may be different than the measured resonance frequency at low temperatures if there are non-equilibrium quasiparticles present that contribute a measurable shift through the kinetic inductance.

The temperature dependent internal quality factor for a 16-μm wide centerline (8-μm wide gap) TiN resonator is presented in Fig. 1(a) with $\langle n\rangle = 10^4$, $Q_c$= 145k, and single-photon $Q_i$ = 400k at base temperature. Similarly, the temperature dependent internal quality factor for a 6-μm wide centerline (3-μm wide gap) Al resonator is shown in Fig. 1(b) with $\langle n\rangle = 10^4$, $Q_c$= 151k, and single-photon $Q_i$ = 412k at base temperature. These two plots also show data (filled circles) and model results for the total $Q_i$ (blue solid line) as well as the calculated contributions from TLSs (red solid line) and quasiparticles (gold solid line). The right axis is used for the effective quasiparticle (black dashed line) and bath (black dotted line) temperatures from the model that describes this data. At base temperature, the power dependence of these resonators shows no saturation even at higher photon numbers indicating that the power-independent loss does not contribute to quality factor measurements reported here. Therefore, $Q_A$ is removed from further consideration.

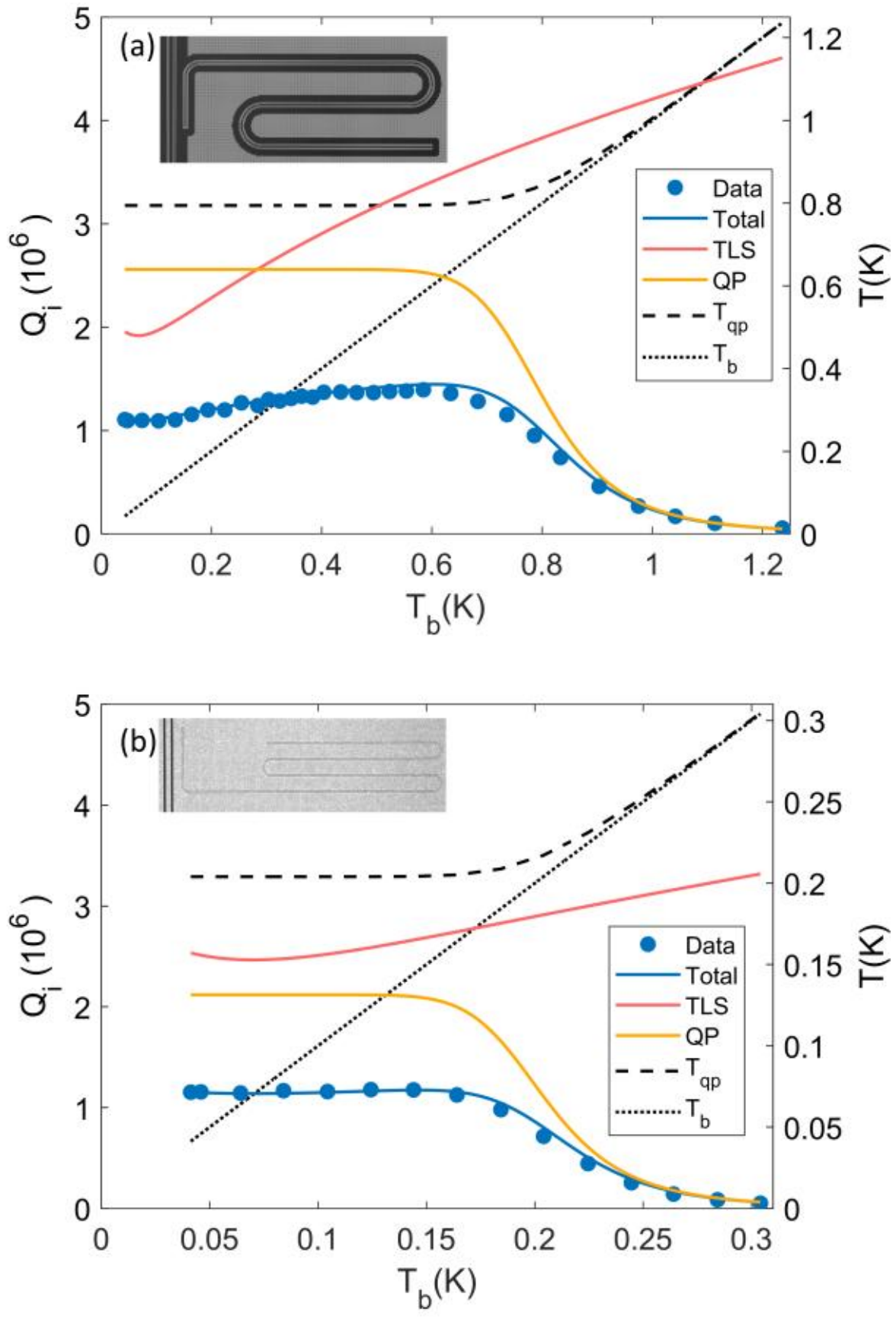

FIG. 1. Comparison of temperature dependent loss for a TiN resonator, R1 (a) and Al resonator, R1 (b) showing data (symbols) and model (blue) results along with contributions from two-level systems (red) and quasiparticles (gold). The measured bath temperature (black dotted line) and the calculated quasiparticle temperatures (black dashed line) are also shown. Inset shows the optical image of the resonators. TiN resonator has 16-μm wide centerline with 8-μm wide gap while Al resonator has 6-μm wide centerline with 3-μm wide gap.

According to these model results, at high bath temperatures, the quasiparticles are thermalized and $T_{qp} = T_b$. At lower temperatures, the quasiparticles are not in equilibrium with the phonons, resulting in $T_{qp} > T_b$. For TiN, the quasiparticle's effective temperature, $T_{qp}$ ~ 794 mK, corresponds to a $n_{qp}$ = 282 μm$^{-3}$. When compared to the BCS Cooper pair number density[31], $n_{cp} = N(0)\Delta(0)$, the relative carrier density ratio is $n_{qp}/n_{cp}$ ~ 1. 19×10$^{-5}$. For Al, $T_{qp}$ = 204 mK, which corresponds to a $n_{qp}$= 103 μm$^{-3}$, and a carrier density ratio $n_{qp}/n_{cp}$ = 3.1 × 10$^{-5}$. It is interesting that the quasiparticle number density is comparable for these two resonators, suggesting that the excess energy is coupled to both samples in a repeatable manner, such as those from an imperfect experimental setup. The greater impact of the quasiparticle loss on the Al resonator is a result of the smaller superconducting gap compared to TiN and is evident through the calculated contributions and relative Q associated with the TLSs and quasiparticles.

## III. RESULTS AND DISCUSSION

The superconducting resonators are measured in transmission mode on the low-temperature stage in an adiabatic diamagnetic refrigerator, configured with 73.2 dB of total attenuation at various thermal plates on the input line. DC blocks are installed at 1 K on both the input and output lines. On the output line, a three-stage rf isolator and low-noise amplifier are installed on the 3 K plate. The transmission spectrum for each resonator is collected by stepping through temperatures and at each temperature measuring $S_{21}$ for a range of different applied powers resulting in the average photon numbers, $\langle n \rangle$, to vary from those corresponding to approximately one photon in the resonator to powers just before the onset of nonlinear effects[41,42]. The internal quality factor is extracted from the measured complex transmission spectra. The power dependent internal quality factor is interpolated for each constant-temperature series to provide data at uniform average photon numbers. Temperature dependence of the quality factors and frequency shifts are calculated using Eqs. (10) & (11). Details related to data collection and interpolation are provided in the supplemental material. The temperature and power dependent internal quality factor and fractional frequency shift data, and model of two TiN resonators are shown in Fig. 2.

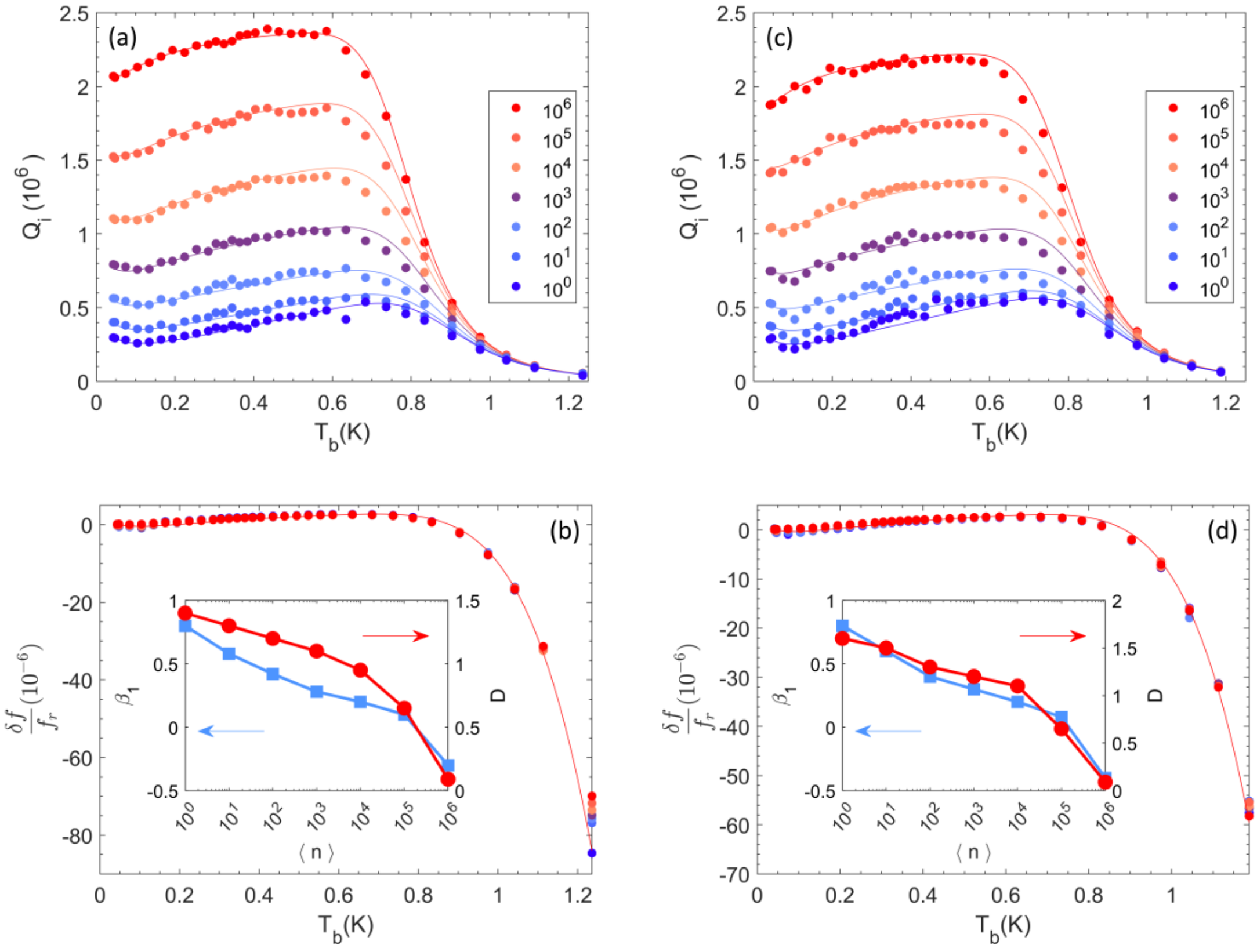

FIG. 2. Temperature and power dependent internal quality factor (a, c) for different $\langle n \rangle$ indicated in the legend, and fractional frequency shift (b, d) data (symbols) and model (lines) of two titanium nitride resonators, R1 (a) & (b), and R2 (c) and (d). The inset graph shows power dependent fitting parameters associated with TLS defects for $\beta_1$ (blue squares) and $D$ (red circles), lines are used to guide the eye.

The internal quality factor in Figs. 2(a) and (c) have three primary temperature regions. In the high-temperature region, losses are dominated by thermally generated equilibrium quasiparticles. The elevated temperatures result in an increased density of thermal quasiparticles where $T_{qp} = T_b$, resulting in a decreasing $Q_i$. For intermediate temperatures, the transition from thermal quasiparticle loss into loss dominated by TLS and non-equilibrium quasiparticles is characterized by the "knee" near 700 mK and the highest values of $Q_i$. The low-temperature regime shows varying behavior depending on the average photon number. Above 100 mK, the typical TLS thermal behavior dominated by $\tanh(\hbar\omega/2k_B T)$ in Eq. (1) is observed. For low powers, there is an increased lift in $Q_i$ at the lowest temperatures, but at high powers, there is a drop which is reflected as a change in sign of $\beta_1$ that indicates a change in TLS-TLS interactions. These trends have been observed in other materials[24,43]. Owing to their empirical nature, D and $\beta_1$ are chosen to be power dependent parameters, which provide the necessary degree of freedom to achieve the fits shown in Figs. 2(a) & (c). Their variation from resonator to resonator depends on the specific arrangement of TLS defects.

The fractional frequency shift shown in Figs. 2(b) & (d) are power independent and the three regimes observed for the internal quality factor are also observed. This behavior is captured by the model. It should be noted, that the number of fit parameters needed to fit the fractional frequency shift is fewer than those needed to fit the internal quality factor. Therefore, the fractional frequency shift is fit first, and those parameters are then used to fit the internal quality factor as described in the supplemental materials.

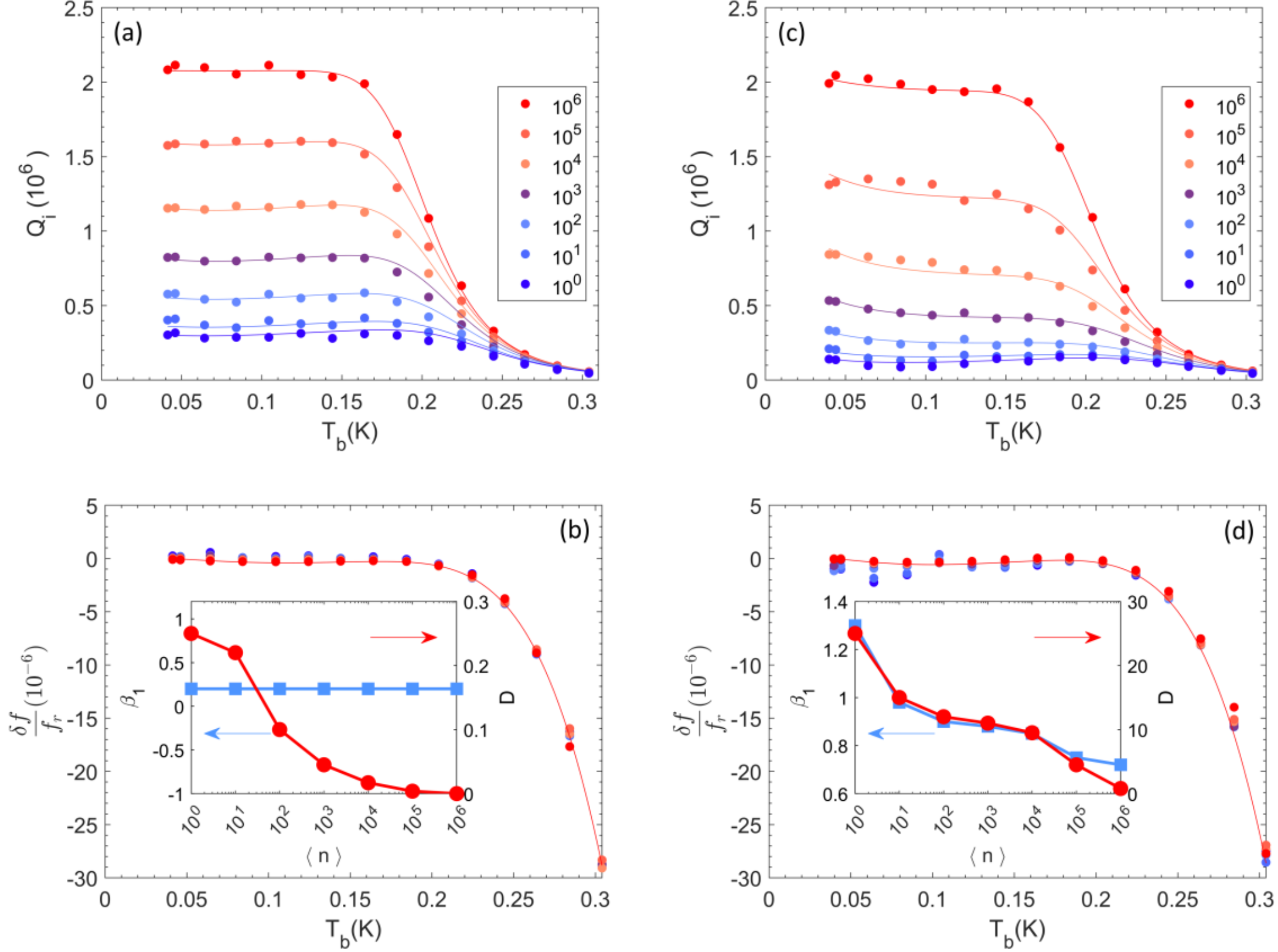

FIG. 3. Temperature and power dependent internal quality factor (a, c) for $\langle n \rangle$ indicated in the legend, and fractional frequency shift (b, d) data (symbols) and model (lines) of two aluminum resonators, R1 (a) & (b), and R2 (c) and (d). The inset graph shows power dependent fitting parameters associated with TLS defects for $\beta_1$ (blue squares) and $D$ (red circles), lines are used to guide the eye.

For the aluminum resonator results shown in Fig. 3, the relative contribution of the quasiparticles and TLS defects in the low-temperature region are evident. The data is flatter in both internal quality factor and fractional frequency shift compared to the TiN resonator that has a larger $T_c$. This flatness is caused by the significant quasiparticle contributions that are either comparable or larger than the TLS contributions at different photon powers. It is important to note that these observations are most likely limited by the thermalization of the experimental setup and may be impacted by incomplete shielding or the short delay between cooling the sample and starting measurements in an adiabatic demagnetization refrigerator that may lead to incomplete cooling of the sample and rf components.

It is useful to acknowledge that by itself, the flat, low-temperature behavior of the R1 $Q_i$ is insufficient to isolate the potential limit imposed by constant loss, $Q_A$, from the nonequilibrium QP loss identified here. Qualitative comparisons of the subtle rounding of the transition ("knee" or "elbow") at mid-temperatures that softens in the presence of quasiparticles is insufficient to isolate the potential impact of $Q_A$. Instead, clarity is provided with this model through the fractional frequency shift that has a flat frequency behavior at low temperatures from quasiparticles, and complex frequency dependence from the TLS defects. Other mechanisms that might be considered such as external flux and radiative/packaging loss would need to modify the complex conductivity to impact both loss and frequency observables in a consistent manner to lower confidence in the assessed model values.

Similar increases in the quality factor at low temperatures that were observed for the TiN resonator have been reported for aluminum resonators[43]. The power dependence of the aluminum resonators in the low and mid temperature regimes is attributed to the power dependence of TLS defects. The comparable contributions of the two sources of loss can cause the power dependent loss assessments to have a lower measure Q, and a weaker power dependence.

All fit parameters for these resonators are reported in Table 1. As evident in both resonators, the model has the most difficulty representing the trends in the intermediate temperature region. It is essential to point out that the extracted values of $I_{ext}$ are not absolute but depend on the choice of $\Gamma_s$. For the four resonators presented here, the lowest calculated temperatures are largely power independent. The critical temperature and resonator frequency are measured parameters, while the rest are fit parameters. Of the parameters that are fit to the model, only the kinetic inductance fraction is directly observable. To qualify the low values of the kinetic inductance, the rf kinetic inductance is estimated from the geometric inductance of coplanar waveguides that do not have trenches. The calculated values for TiN is 0.93 pH/☐, and 0.31 pH/☐ for Al. These values are relatively comparable to the kinetic inductance determined from dc transport measurements and BCS theory, to be 1.9 pH/☐ for TiN, and 0.085 pH/☐ for Al. Additional details are in the supplemental materials.

Table 1: Comparison of TLS loss parameters determined by a power dependent TLS-only model and power-temperature dependent TLS-quasiparticle model for this TiN resonator. Parameters marked with * are determined by experiments.

| **Parameter** | **TiN R1** | **TiN R2** | **Al R1** | **Al R2** |
|---|---|---|---|---|
| $Q_{TLS0}$ | 0.11 | 0.11 | 0.12 | 0.08 |
| $D$ | {1.4, 1.3, 1.2, 1.1, 0.95, 0.65, 0.09} | {1.6 1.5 1.3 1.2 1.1 0.65 0.09} | {0.25, 0.22, 0.1, 0.045, 0.017, 0.0038, 2e-5} | {25, 15, 12, 11, 9.5, 4.5, 0.80} |
| $\beta_1$ | {0.8, 0.58, 0.42, 0.28, 0.2, 0.1, -0.3} | {0.8, 0.6, 0.4, 0.30, 0.20, 0.08, -0.4} | 0.2 | {1.3, 0.98, 0.90, 0.88, 0.85, 0.75, 0.72} |
| $\beta_2$ | 0.55 | 0.55 | 0.15 | 0.57 |
| $I_{ext}$ ($\mu m^{-3}s^{-1}$) | $2.37\times10^{8}$ | $2.75\times10^{8}$ | $3.26\times10^{7}$ | $2.97\times10^{7}$ |
| $\Gamma_S$ (MHz) | 0.82 | 0.82 | 0.31 | 0.31 |
| $\alpha_{L_{ki}}$ | 0.123 | 0.123 | 0.108 | 0.108 |
| $f_r$ (GHz)* | 5.0186 | 4.4695 | 5.2508 | 4.7371 |
| $T_C$ (K)* | 5.29 | 5.29 | 1.25 | 1.25 |

The empirical description of TLS defect loss adopts the form from ref [24]. Of interest is the relationship between the internal quality factor and potential trends in the relative TLS decoherence time, $T_{2_{TLS}} \propto \left(DT^{\beta_1}\right)^{-1}$. The temperature dependent trends of the TLS decoherence time are shown in Fig. 4.

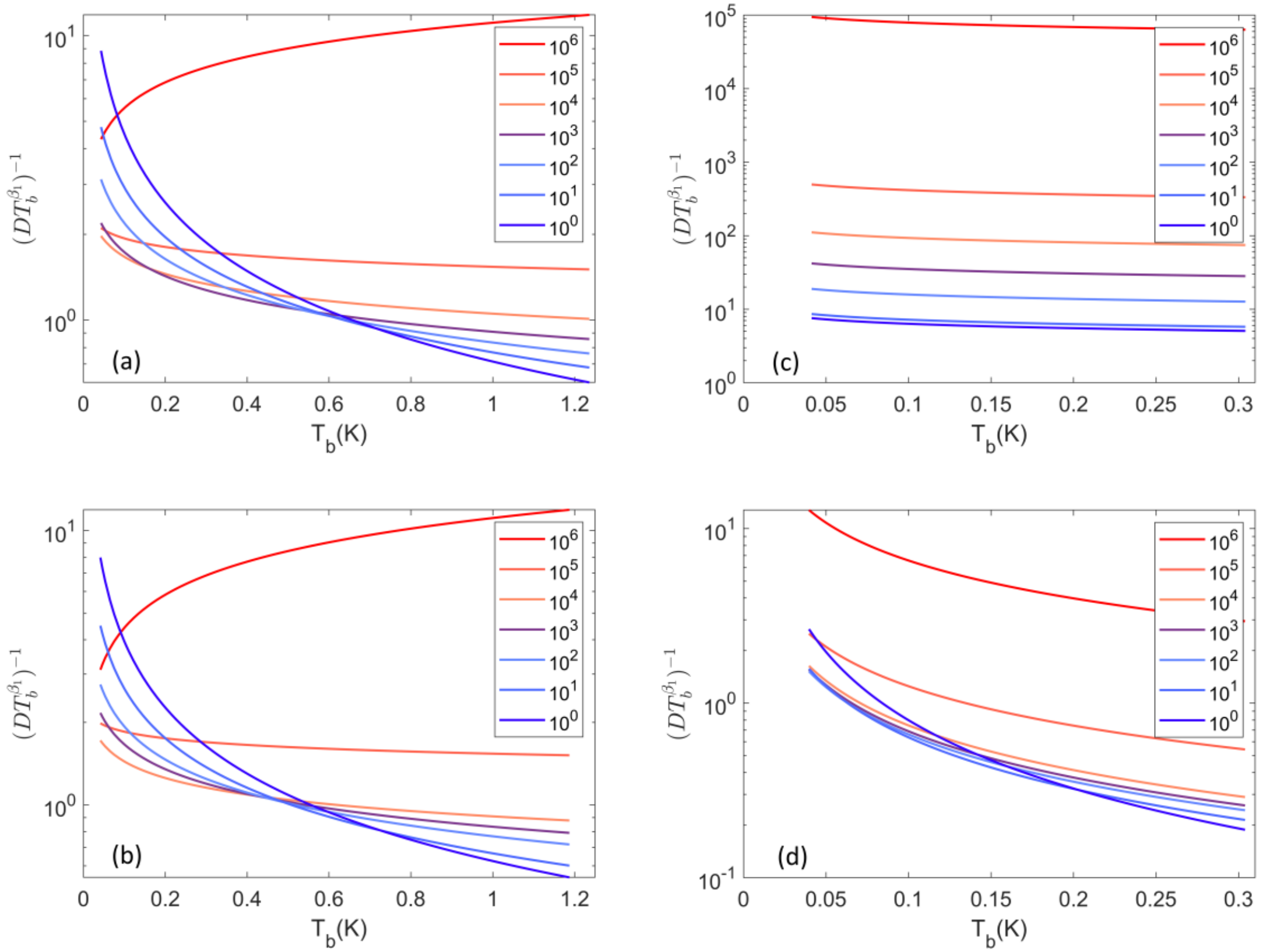

FIG. 4. Comparison of the temperature and power dependent TLS decoherence time ($T_{2_{TLS}} \propto \left(DT^{\beta_1}\right)^{-1}$) for the TiN resonators R1 (a) and R2 (b), and the Al resonators R1 (c) and R2 (d) for different average photon numbers as shown in the legend.

There are three key observations from these two graphs. The first is a comparison between resonator materials and the relative contribution of quasiparticles that tend to flatten out the temperature dependence. This is observed for high temperatures for the resonators in Fig 4(a), (b), and (d), and all temperatures and powers for the Al resonator R1 in Fig 4(c). The key being the different $T_C$ of these two materials and the relatively high $I_{ext}$ from the experimental setup. The general trend of decreasing decoherence time for increasing temperature is consistent with previous observations[44].

The second observation is that the TLS decoherence time first decreases and then increases with power at lower temperatures for TiN. This behavior might be the result of the TLS population being in the ground state with a smaller interaction strength at low powers resulting in a relatively high TLS decoherence time, that is complimented by a region of high decoherence time when the TLS population is becoming polarized. It is also possible that athermal quasiparticles contribute to these trends.

The third observation is the change in TLS decoherence time at the highest temperatures where the TLS decoherence time increases with power. In this region, it is possible that the TLS bath is becoming saturated and polarized as applied power is increased. These effects may also result from athermal quasiparticles where increases in $Q_i$ have been observed in quasiparticle-limited systems[20,45]. To better understand this behavior, a completely different model that describes the microscopic behavior of non-thermal quasiparticle dynamics is needed.

The results and analysis presented here is indeed complex. There are over a hundred measurements reported on each resonator under different temperatures and powers. There are many free parameters, so a structured fitting process is used to work through the parameters to eliminate arbitrary numerical results. First, using alternative experiments to determine or baseline observable parameters such as $T_c$ and $f_r$, as well as literature values for fundamental materials parameters such as $\Gamma_R$ are used to reduce the number of parameters. Next, the fractional frequency shift is fit because it has fewer parameters than the internal quality factor enabling the determination of $\alpha_{L_{ki}}$, and $Q^0_{TLS}$. Only then is the quality factor fit to determine the remaining parameters, using the high-power results to refine $\Gamma_S$. For the results presented here, there is no constant loss considered, $Q_A \gg Q_{QP}, Q_{TLS}$. A key feature of this model and process is that there is consistency of the quasiparticle parameters across resonators on a single chip as expected from a homogeneous material.

It is common for superconducting resonators to be measured under a smaller range of conditions such as low-temperature power-sweeps, or high-power temperature-sweeps to maximize measurement efficiency. The results of such measurements are often analyzed with parts of the model presented here. The successful application of the model on all the data collected from the Al and TiN resonators with a single set of parameters across loss and frequency shift demonstrates the utility of the complete model allowing aspects of the TLS-TLS behavior to be more holistically assessed.

## IV. Conclusion

In conclusion, understanding, disentangling, and mitigating the losses from two-level systems and quasiparticles is an important topic for quantum information science. The power and temperature dependent microwave loss and frequency shift model presented describes the loss and frequency shift from two resonators made from different materials over a wide range of experimental conditions. The small set of parameters needed to describe loss and frequency shift across all measured conditions has value to extract parameters that represent the sample, helps reduce misinterpreting two-level system and quasiparticle contributions, and may help characterize TLS-TLS interactions. These results may help data analysis on various experimental configurations and assist in determining strategic needs for improving either device fabrication or measurement.

# Power and temperature dependent model for High Q superconductor resonators

# Supplemental Material

A. Alexander[1, 2], C.G. Weddle[1], C.J.K. Richardson[1, 3]

[1] Laboratory for Physical Sciences, University of Maryland, 8050 Greenmead Dr., College Park MD 20740, USA

[2] Department of Physics, University of Maryland, College Park, Maryland 20742, USA

[3] Department of Material Science and Engineering, University of Maryland, College Park, Maryland 20742, USA

## S1. EXPERIMENTAL DETAILS

The two films reported here are grown on 3-inch, high-resistivity ($\rho > 5$ kΩ-cm), float zone refined silicon (111) wafers in a DCA M600 plasma-assisted molecular beam epitaxy (MBE) system with a base pressure less than $10^{-10}$ Torr. Both titanium and aluminum are evaporated from effusion cells. Before growth, wafers are cleaned by a solvent series of acetone, methanol, and isopropyl alcohol to remove organic contaminants. Then sequentially etched in 5% HF and 40% $NH_4F$ solutions to remove the native surface oxide and obtain step edges. In the UHV chamber, the wafers are heated to 800 °C to remove surface hydrogen, fluorine, and residual oxide.

The 100-nm-thick titanium nitride film is grown by the near-simultaneous introduction of titanium and nitrogen flux. The film is grown with a substrate temperature measured by thermocouple of 800 °C under nitrogen-rich conditions at a rate of 0.003 nm/s. Following growth, the wafer is cooled under nitrogen overpressure.

100-nm-thick aluminum films are grown on a nitride surface. Following UHV heating to thermally clean the surface, the wafer is exposed to a nitrogen flux from the plasma cell for 15 min at 800 °C. The substrate is cooled to 100 °C before aluminum growth at a rate of 0.032 nm/s. The symmetric X-ray diffraction shows the presence of relaxed aluminum and titanium nitride films. The critical temperature ($T_C$) is measured with DC resistance measurements of unpatterned films using 4-point measurements. The $T_C$ for the titanium nitride film is 5.3 K, and 1.2 K for the aluminum film. Additional information is available in REFs [1–3].

The resonator chips are fabricated in a 5x stepper. For the Al sample, the pattern defined by positive i-line photoresist (OiR 906-10) is wet etched with Transene Type A aluminum etchant to define the resonator pattern. For the TiN sample, the pattern using i-line negative photoresist (AZ® nLOF 2020) is dry etched in a reactive ion etch inductively coupled plasma system using a gas mixture of 12.5 sccm $BCl_3$ and 2.5 sccm $Cl_2$ and RF power of 500 W. There is no measurable trench etched into the Al sample, while the TiN sample has a 300-nm coplanar waveguide trench[4]. Resist is removed from the samples using a series of Microposit 1165 solvent baths and a final clean with deionized water and isopropyl alcohol.

Resonators are measured in a High Precision Devices (HPD) adiabatic demagnetization refrigerator (ADR). The 0-dBm input signal from the vector network analyzer feeds into a variable attenuator. The input line contains a series of attenuators at different temperature stages (3 dB at 50 K, 30 dB at 3 K, 20 dB at 1 K) and a DC block at 1 K to reduce thermal blackbody energy from exciting the device. The small diameter stainless-steel semi-rigid coaxial microwave lines used below the 3K plate introduce an additional 20 dB of attenuation for both the input and output lines. Thermal protection of the output line is accomplished with a DC block at 1 K, and a microwave isolated HEMT low noise amplifier at 3K. In the low-temperature stage, the resonator device is wire bonded into a custom copper package in a light-tight copper radiation shield and Mu-metal magnetic shield.[5]. The schematic of the refrigerator and measurement configuration is shown in Fig. S1.

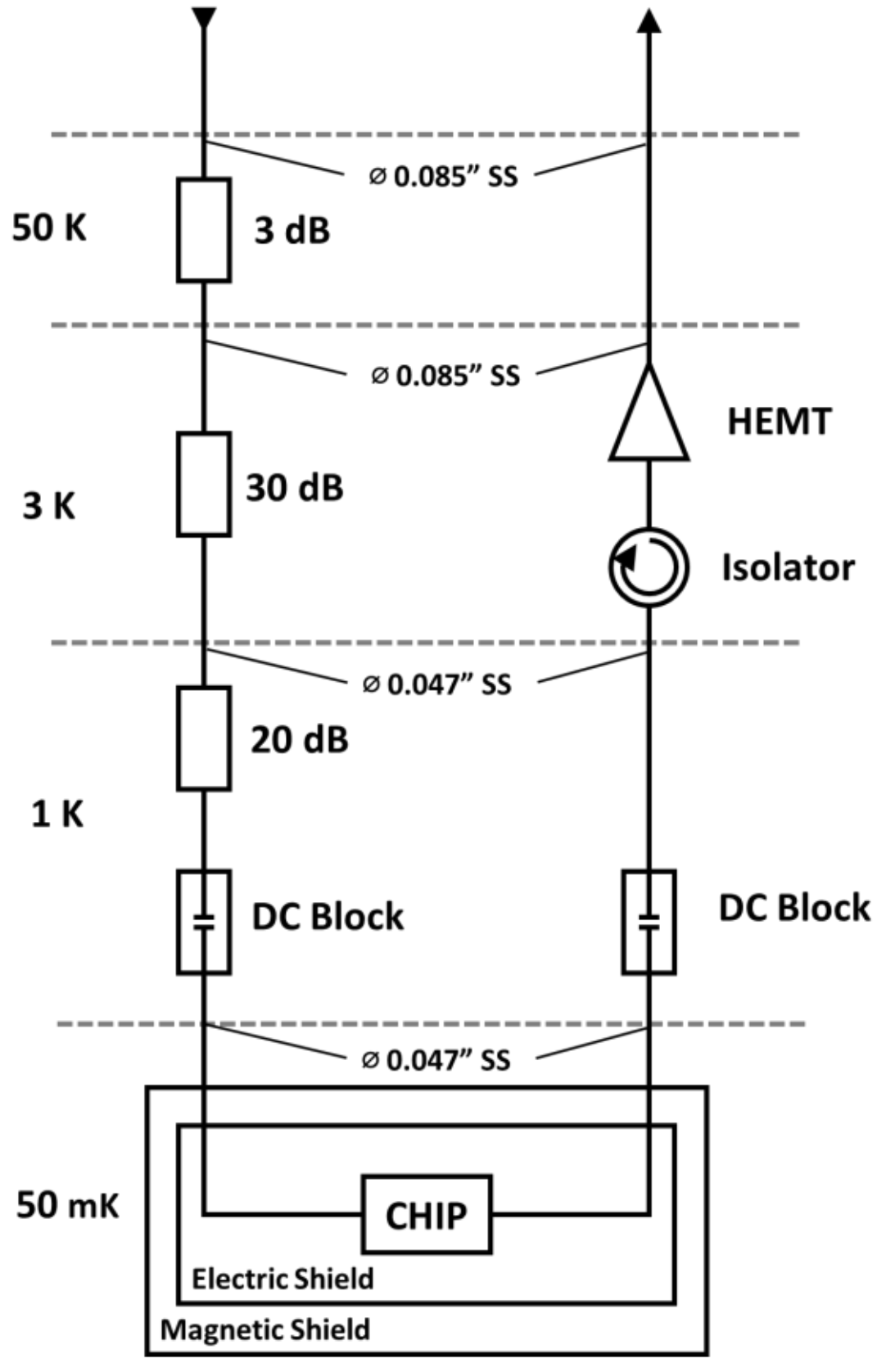

FIG. S1. Schematic of the rf measurement and shielding configuration used for resonator measurements in the adiabatic demagnification refrigerator.

## S2. RESONATOR MEASUREMENT AND FIT

The coplanar quarter-wave resonators are capacitively coupled to a transmission line bisecting the chip. The complex S-parameter transmission matrix element, $S_{21}(f)$, exhibits a dip in the transmitted power around the center frequency, $f_r$. The normalized transmission coefficient of the resonator, including all the linear non-idealities, can be written as[2]

$$S_{21}(f) = e^{i(\theta+\kappa L)}\left[1 - \frac{Q_t}{Q_c}\frac{e^{i\phi}}{1 + 2iQ_T\left(\frac{f - f_r}{f_r}\right)}\right]. \tag{S1}$$

In all, there are seven parameters that need to be extracted from fitting the data with the fit function. The global phase of the measurement, $\theta$, the uncompensated rf pathlength, $\kappa L$ with $\kappa = 2\pi/\lambda$ being the wavenumber of the rf signal ($\lambda$ is the guided rf wavelength), and the coupling phase, $\phi$. The loaded quality factor, $Q_t$, includes the energy loss via coupling between the transmission line and resonator, $Q_c$, and internal loss, $\tan\delta = Q_i^{-1}$,

$$\frac{1}{Q_t} = \frac{1}{Q_i} + \frac{1}{Q_c}. \tag{S2}$$

where the effective coupling quality factor, $Q_c^* = Q_c e^{-i\phi}$.

Multiple perspectives of the data are useful to understand nonideal features of the data. The ideal parametric plot of normalized $\tilde{S}_{21}$ data in the complex plane should form a circle centered on the x-axis. Coupling phase can lead to circles rotated around endpoints far from the resonance frequency and imperfect path length compensation can lead to a spiral shape. Additional complications can arise from an indirect determination of $Q_i$ that is not directly determined from Eq (15). Instead, the normalized inverse $S_{21}$ gives $Q_i$ directly as a fitting parameter.

$$S_{21}^{-1} = e^{-i(\theta+\kappa L)}\left[1 + \frac{Q_i}{Q_c}\,\frac{e^{i\phi}}{1 + 2iQ_i\left(\frac{f - f_r}{f_r}\right)}\right]. \tag{S3}$$

In Fig. S2, the clear differences between dips $S_{21}$ versus peaks in $S_{21}^{-1}$ are evident, it should be noted that the spacing of data on the parametric plots are significantly different. This leads to different data collection requirements of the spectral density of data needed to adequately use both analysis equations.

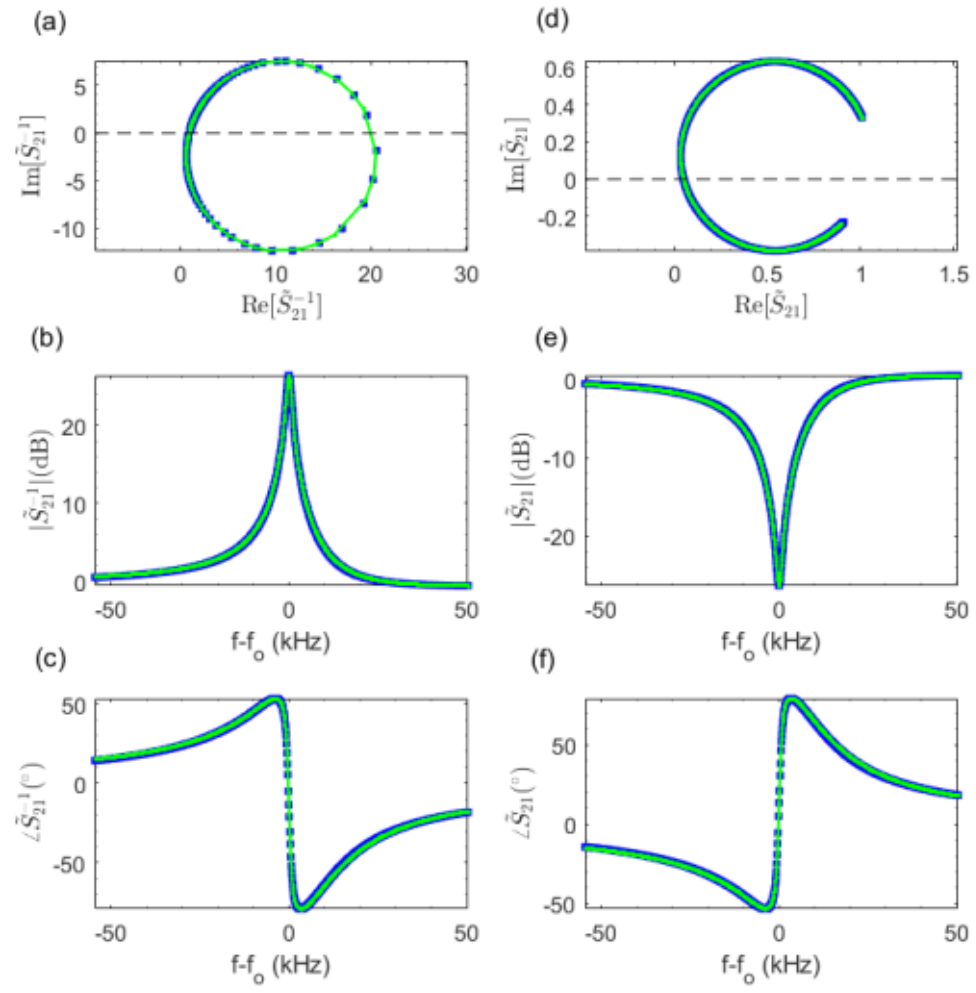

FIG. S2. Representative normalized $S_{21}$ parametric form (a), magnitude (b), and phase (c) signal from a resonator. Comparable plots using the same data showing $S_{21}^{-1}$ parametric form (d), magnitude (e), and phase (f). Measured data is first normalized, and corrected for the linear phase artifact that is most clearly evident as a spiral character in the parametric representation of $S_{21}$. The remaining resonator parameters are extracted from fitting the complex representation of $S_{21}^{-1}$. This approach enables the robust fitting of the observed resonances in the presence of measurement artifacts that, if not properly mitigated, can cause fitting errors that are most evident in the scatter of the power and temperature dependent measurements.

For each resonator, power dependent data is collected at different temperatures using the variable attenuator and fixed output from the rf source. The smallest applied power at the device is -153 dBm. After collecting the power dependent data at a specific temperature, the ADR temperature is increased by regulating the temperature through the current passing through the magnet surrounding the ADR salts. The temperature is adjusted from a base temperature of 50 mK to a maximum temperature of $T_C/4$. The maximum temperature measured for Al resonators is 320 mK, and 1.2 K for TiN resonators.

The resonator data is fit to extract the internal quality factor and the average photon number, $\langle n_p \rangle$, in the resonator is calculated from the input power ($P_{in}$)[6],

$$\langle n_p \rangle = \frac{2}{\hbar\omega_r}\frac{Q_T^2}{Q_C}P_{in}, \tag{S4}$$

where $\omega_r = 2\pi f_r$ is the angular resonance frequency. Even though data is collected using regularly spaced values of the applied power the resulting average photon number is not uniformly spaced because of the specific total quality factor of each resonator that is power dependent. To facilitate analysis of the power dependent results data is interpolated. The standard power dependent TLS model[1,2] is applied to power dependent $Q_I$ data,

$$\frac{1}{Q_I} = \frac{1}{Q_A} + \frac{1}{Q_{TLS}^{T_b}} \frac{1}{\sqrt{1 + \left(\frac{n_p}{n_c}\right)^{\alpha}}}. \quad (S5)$$

Here $Q_A$, $Q_{TLS}^{T_b}$, $n_c$ and $\alpha$ are free parameters. Using the fit functions, $Q_I$ is interpolated for the photon numbers $\{10^0, 10^1, 10^2, 10^3, 10^4, 10^5, 10^6\}$. Surprisingly, this fit is excellent for all measured data. An exemplar is shown in Fig. S3, showing the power dependent data of an Al resonator measured at 45 mK.

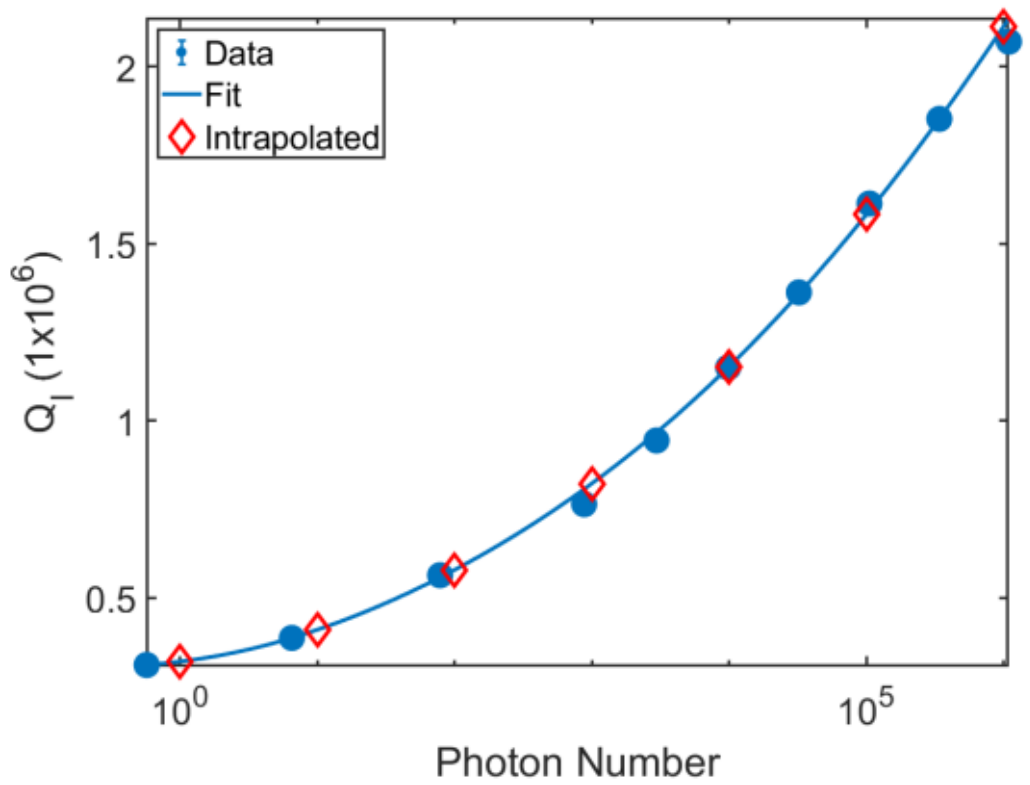

FIG. S3. Power-dependent interpolation of the internal quality factor from an Al resonator was measured at 45 mK. The internal quality factor determined from the measured data (blue circles), power dependent fit (line), and interpolated values (red asterisks) provides consistent representations of the measured data.

## S3. SUPERCONDUCTING PARAMETERS

The two-temperature model has several material parameters that are determined from published literature values, experimental measurements, and free parameters. Details about how each parameter is determined are cataloged in Table S1 and detailed in this section.

Table S1: Material parameters for Al and TiN

| Parameter | TiN | Al |
|---|---|---|
| $\xi_o$(nm) | 105[7] | 1729[8] |
| $\lambda_{LO}$(nm) | 575[9] | 15.4[8] |
| $\tau_o$(ns) | 5.5[10] | 110[11] |
| $N(0)(\mu m^{-3} eV^{-1})$ | $2.96\times 10^{10}$ [10] | $1.72 \times 10^{10}$ [12] |
| $\Gamma_R$ ($\mu m^3$/s) | 83 | 28 |
| $l$ (nm) | 100 | 100 |
| $\gamma$ | -1 | -1/3 |
| $T_C$ (K) | 5.29 | 1.26 |
| $L_{k/\square}$ (pH/□) | 1.90 | 0.09 |
| $R_s$ ( Ω/□) | 7.30 | 0.08 |
| RRR | 2.33 | 24 |

**Quasiparticle recombination rate, $\Gamma_R$**

The quasiparticle recombination rate has been studied for both Al and TiN materials. For the temperature range $T_c/10 < T < T_c/3$, the quasiparticle recombination rate $\Gamma_R$ is approximated as[11]

$$\Gamma_R = 2\left(\frac{\Delta(0)}{k_B T_C}\right)^3 \left(\frac{1}{N(0)\Delta(0)\tau_o}\right). \tag{S6}$$

Where, $\Delta(0)$ is the BCS superconducting gap parameter at T = 0 K, $T_C$ is the superconducting critical transition temperature, $N(0)$ is the normal state single-spin density[13], and $\tau_o$ is a characteristic time constant describing material-dependent electron-phonon interaction[10,11,14]. The calculated recombination rate, and literature values for the normal state spin density and characteristic time constant are cataloged in Table S1.

**Quasiparticle relaxation rate, $\Gamma_S$**

The quasiparticle relaxation rate has not been independently determined, and its value is most likely dependent on the degree to which the quasiparticle distribution is athermal. The two-temperature model described here cannot adequately capture the complex athermal behavior that is likely possible in different experimental conditions. However, as an approximate description of the potential excess energy in the quasiparticles, the relaxation rate is captured in a bounded fit parameter.

The quasiparticle relaxation rate can be in the low-temperature limit, $n_{qp}(T_{qp}) \gg n_{qp}(T_b)$, where Eq. (5) can be reduced to

$$\Gamma_s = \frac{I_{ext} - \Gamma_R n_{qp}(T_{min})^2}{n_{qp}(T_{min})}, \tag{S7}$$

Using this relationship to calculate $\Gamma_s$ for several values of $I_{ext}$ results in the various fits shown in Figure S4 for the Qi corresponding to the highest measured photon number of $\langle n_p \rangle = 10^6$.

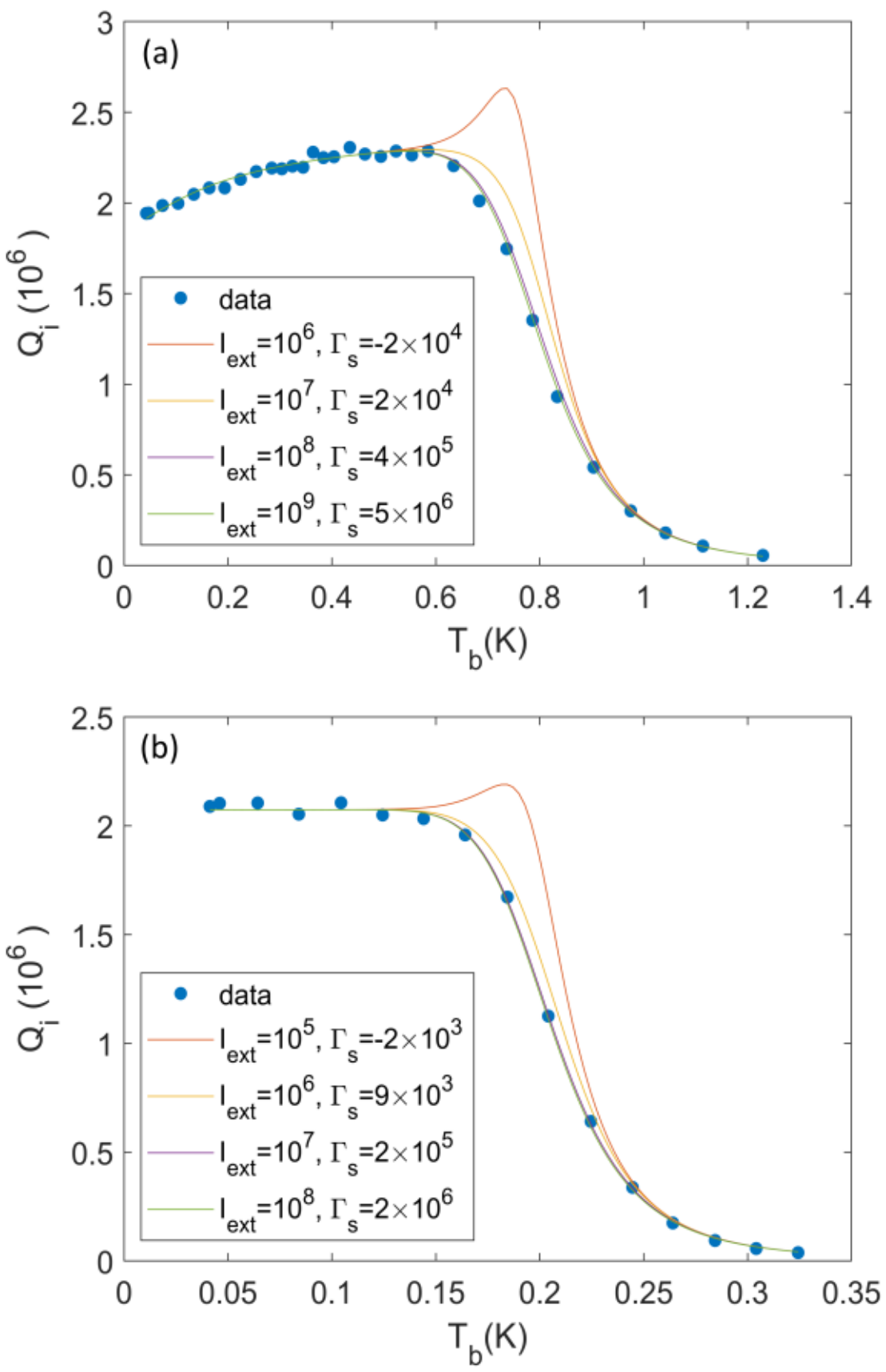

FIG. S4. Temperature dependent quality factors at $\langle n_p \rangle = 10^6$ with various fits with different values of $\mathrm{I_{ext}}$ and $\Gamma_\mathrm{s}$ associated values that are calculated using Eq. (S7) for (a) TiN and (b) Al. The $\mathrm{I_{ext}}$ is expressed in $\mu\mathrm{m}^{-3}\mathrm{s}^{-1}$ and $\Gamma_\mathrm{s}$ is expressed in MHz.

Under the constraints imposed by Eq. (S7), there is a convergence of curves above some value of $\mathrm{I_{ext}}$ and the corresponding $\Gamma_\mathrm{s}$ that does not result in changes at low or high temperatures. It is apparent from this exercise that the impact of the single particle relaxation is most critical at intermediate temperatures. The lowest value of $\mathrm{I_{ext}}$ that results in a consistent fit is used as the parameter. The fit values of the relaxation rate and external excitation density that are determined from the analyzed resonator data are cataloged in Table 1 in the primary article. The variations due to different values of $\mathrm{I_{ext}}$ and its corresponding $\Gamma_s$ are not so dominant in the fractional frequency shift.

**Characteristic lengths, γ**

The particular value of γ in the Eqs. (10) and (11), depends on the superconducting parameters of the thin film material. In particular, the value is determined by the relative values of the London penetration depth, $\lambda_{\mathrm{LO}}$, mean free path, l, and coherence length, $\xi_0$, in the usual manner,

$$\gamma = \begin{cases} -\frac{1}{3}, & \xi_0 \gg \lambda_{\mathrm{LO}} \;\&\; \mathrm{l} \gg \lambda_{\mathrm{LO}} \quad \text{(Extreme anomalous thick film limit)} \\ -\frac{1}{2}, & \mathrm{l} \ll \xi_0 \;\&\; \mathrm{l} \gg \lambda_{\mathrm{LO}} \quad \text{(Dirty limit)} \\ -1, & \mathrm{l} \ll \xi_0 \;\&\; \mathrm{l} \ll \lambda_{\mathrm{LO}} \quad \text{(Thin film limit)} \end{cases}, \tag{S8}$$

For thin samples such as those explored here, the film thickness determines the mean free path. The parameters in Table S1 categorize Al as being in the extreme anomalous thick film limit, while TiN is in the thin film.

**Superconducting Critical Temperature, $\mathrm{T_C}$**

The critical temperature, $T_C$, is determined from DC conductivity measurements in a van der Pauw geometry. This configuration with contacts on the four corners of a square sample is suitable for homogeneous films and allows the sheet resistance, $R_s$, to be determined from the measured resistance, R, through $R_s = \pi R/\ln(2)$ . The measured critical temperature of the TiN film is 5.3 K, with a normal state sheet resistance of 7.2985 Ω /• , and residual resistance ratio, RRR = 2.33. The measured critical temperature for aluminum is 1.3 K, with $R_s = 0.0771$ Ω /• , and a RRR = 24.

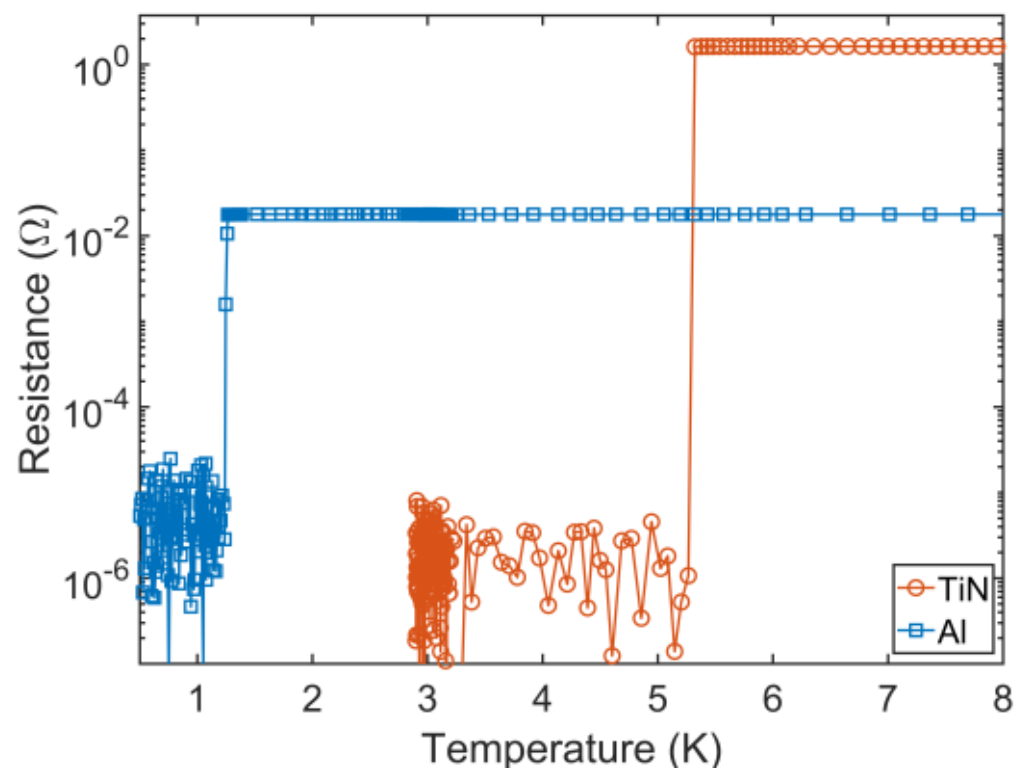

FIG. S5. DC conductivity measurement for the Al and TiN film

**Kinetic Inductance Fraction, $\alpha_{L_{ki}}$**

The kinetic inductance can be determined by two methods. At rf frequencies, the resonant frequency can be used to determine the fraction of the total inductance that is attributed to either the material or waveguide geometry, and through DC conductivity measurements interpreted through BCS theory. From DC measurements, the sheet kinetic inductance of a superconductor is given by[15]

$$L_{k/\square} = \frac{R_s \hbar}{\pi \Delta(0)} \frac{1}{\tanh\left(\frac{\Delta(0)}{2k_B T}\right)}. \tag{S9}$$

Where, $R_s$, is the normal state sheet resistance just above the critical temperature. Using the values from Fig S5, the calculated $L_{k/\square}$ for TiN is 1.9 pH/☐, and 0.085 pH/☐ for Al.

From the design of the resonator structure, the geometric inductance per unit length of a zero-thickness coplanar waveguide can be calculated using the conformal mapping technique,[16]

$$L_g = \frac{\mu_o K(k')}{4K(k)}. \tag{S10}$$

Where $\mu_o$is the permeability of free space, $k = w/(w + 2g),$ is the ratio of the central conductor width, w, and the separation between the ground planes, g, K(k) is the complete elliptic integral of the first kind, and $k' = \sqrt{1 - k^2}$. The kinetic inductance fraction, $\alpha_{L_{ki}}$, is calculated from the two-temperature model in the main text, the rf kinetic inductance is calculated using the relation

$$L_{k/\square}^{rf} = \frac{\alpha_{L_{ki}}}{1 - \alpha_{L_{ki}}} L_g w. \tag{S11}$$

The rf kinetic inductance determined for TiN is 0.93 pH/☐, and 0.31 pH/☐ for Al.

There are often discrepancies between the low frequency and rf kinetic inductances. One particular source of error is the impact of trenching in the coplanar waveguide gap that changes the geometry and geometric inductance from this simple assessment described by S10.

**Resonator frequency, $f_r$**

It is noteworthy that in the two-temperature model the existence of an elevated quasiparticle temperature, and the non-zero kinetic inductance will result in a measured resonator frequency that is slightly different than that arising from the geometry of the guided mode. Even the resonator frequency measured at the lowest power and the lowest temperature has some non-zero TLS and quasiparticle contribution. A frequency offset is needed to account for these contributions when determining the sample parameters in the presented model.